%% file: SKA_revision-final.tex
\documentclass[a4paper,11pt]{article}
\usepackage{aaskaiid}
\usepackage{orcidlink}

\title{Identification and Study of Irregular Radio Sources with SKA Continuum Surveys}
\ShortTitle{Irregular Radio Galaxies}

\author[1,5]{Tapan K. Sasmal\orcidlink{0000-0001-9251-9456}}
\ShortName{Sasmal et al.}
\author[1]{Xuelei Chen\orcidlink{0000-0001-6475-8863}}
\author[2]{Baoqiang Lao\orcidlink{0000-0002-3426-3269}}
\author[3]{Soumen Kumar Bera\orcidlink{0000-0001-7121-4258}}
\author[1]{Yougang Wang\orcidlink{0000-0003-0631-568X}}
\author[4]{Soumen Mondal}
\author[3]{Taotao Fang\orcidlink{0000-0002-2853-3808}}
\author[5,6]{Renyue Cen\orcidlink{0000-0001-8531-9536}}

\affiliation[1]{National Astronomical Observatories, Chinese Academy of Sciences, Beijing 100101, People’s Republic of China}
\emailAdd{tapan.phys@gmail.com}
\emailAdd{xuelei@bao.ac.cn}
\affiliation[2]{Guangxi Key Laboratory of Brain-inspired Computing and Intelligent Chips, School of Electronic and Information Engineering, Guangxi Normal University, Guilin 541004, People’s Republic of China}
\affiliation[3]{Department of Astronomy, Xiamen University, Xiamen, Fujian 361005, People’s Republic of China}
\affiliation[4]{Department of Physics, Jadavpur University, Kolkata 700032, India}
\affiliation[5]{Center for Cosmology and Computational Astrophysics, Institute for Advanced Study in Physics, Zhejiang University, Hangzhou, 310058, People’s Republic of China}
\affiliation[6]{Institute for Astronomy, School of Physics, Zhejiang University, Hangzhou, 310058, People’s Republic of China}
\emailAdd{renyuecen@zju.edu.cn}

\abstract{Radio galaxies show a wide range of morphologies, from regular double-lobed systems to more complex and distorted radio structures. In this chapter, we focus on irregular radio morphologies, defined as sources in which the radio jets and lobes deviate from a straight and symmetric structure. Bent-tail radio galaxies and winged radio galaxies are two important examples of such sources. Bent-tail radio galaxies show curved jets or lobes, mainly shaped by the interaction between radio plasma and the dense intracluster or intragroup medium. Winged radio galaxies show faint off-axis emission, which may be related to plasma backflow, jet reorientation, episodic activity, galaxy mergers, or environmental asymmetry. The Square Kilometre Array (SKA) continuum surveys will provide the sensitivity, angular resolution, frequency coverage, and image quality required to identify and study large samples of such irregular radio galaxies. These data will make it possible to detect faint extended structures, including diffuse tails, weak bridges, remnant lobes, and low-surface-brightness wings. The identification and classification of these sources will require a combination of machine-learning methods, quantitative morphology measurements, multi-wavelength host-galaxy association, and expert visual inspection. The study of irregular radio galaxies with SKA data will help to connect radio morphology with host-galaxy properties, Active Galactic Nucleus (AGN) activity, jet power, and surrounding environment. Such studies will provide important insight into jet--environment interactions, AGN feedback, the dynamical state of galaxy groups and clusters, and the evolution of radio galaxies across cosmic time.}

\begin{document}
\maketitle

\section{Introduction}

Understanding the radio morphology of extragalactic radio sources remains a fundamental challenge in astrophysics, as it encodes key information about the evolution of active galactic nuclei (AGN), jet dynamics, and interactions with the surrounding environment. Historically, radio galaxies have been classified into the Fanaroff--Riley (FR) types, where type I (FR-I) sources exhibit edge-darkened jets; on the other hand, type II (FR-II) sources display edge-brightened lobes associated with powerful, collimated outflows \citep{Fanaroff1974}. While this canonical framework successfully explains the large-scale structure of many radio galaxies, recent deep and high-resolution surveys have revealed a significant population of sources with complex and unusual morphologies that deviate from standard models. These include distorted, asymmetric, and multi-component structures that likely arise from environmental effects such as interactions with the intracluster medium, galaxy motions, or episodic AGN activity. Studying these irregular morphologies is therefore essential for advancing our understanding of jet--environment coupling, feedback processes, and the life cycles of radio galaxies beyond the traditional FR classification scheme.

In this chapter, the term `irregular radio morphology' refers to radio sources in which the radio jets and lobes deviate from a straight, symmetric structure. Among these classes, bent-tail (BT) radio galaxies are among the important populations for studying the interaction between radio jets and dense environments. These sources exhibit radio jets and lobes that deviate from a linear morphology, producing characteristic C-, L-, V-, or U-shaped structures \citep{Ryle1968, Blanton2000, Sasmal2022}. Depending on the opening angle between the two jets, BT radio galaxies are generally classified into narrow-angle tail (NAT) and wide-angle tail (WAT) sources. NAT galaxies typically show strongly bent jets with small opening angles, whereas WAT galaxies possess comparatively wider jet structures \citep{Owen1976}. The distorted morphology of BT sources is widely believed to arise from hydrodynamic interactions between relativistic jets and the surrounding intracluster medium (ICM), in which ram pressure from the host galaxy's motion through the dense cluster environment deflects the jets away from their original direction \citep{Begelman1979}. Because of their strong environmental dependence, BT radio galaxies are considered reliable tracers of galaxy clusters and dynamically active large-scale environments, particularly at intermediate and high redshifts \citep{Dehghan2014, Mguda2015}. Large-area radio continuum surveys conducted with Low-Frequency Array (LOFAR), Australian Square Kilometre Array Pathfinder (ASKAP), and the  Very Large Array (VLA) have significantly increased the number of known BT sources in recent years, enabling statistical investigations of jet bending, cluster weather, and AGN--environment interactions \citep{Sasmal2022, Pal2023, Lao2025}.

\begin{figure*}
\vbox{
\centerline{
\includegraphics[angle=0,width=7.0cm,height=7cm,trim={0 0cm 0 0},clip]{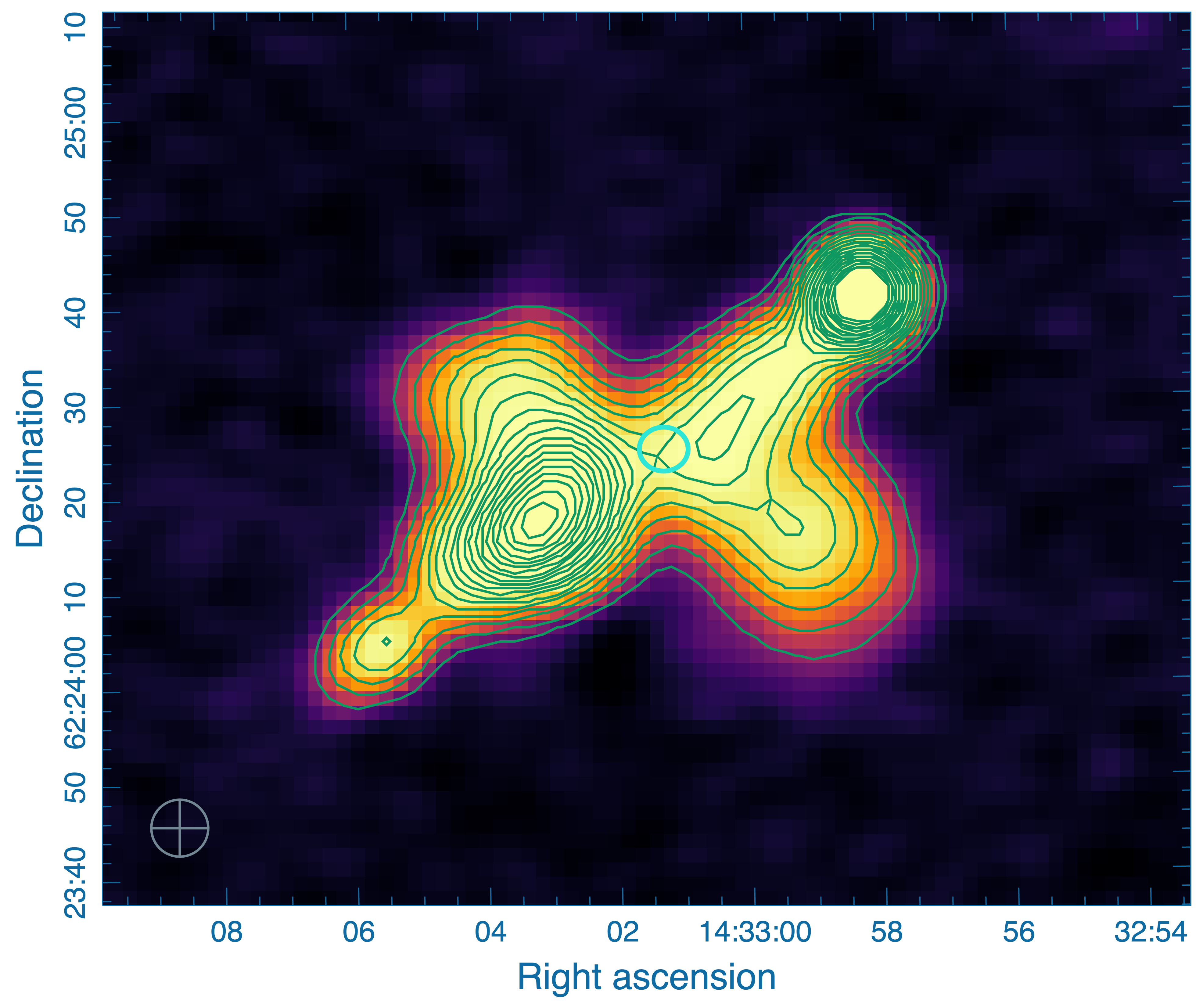}
\includegraphics[angle=0,width=7.0cm,height=7cm,trim={0 0cm 0 0},clip]{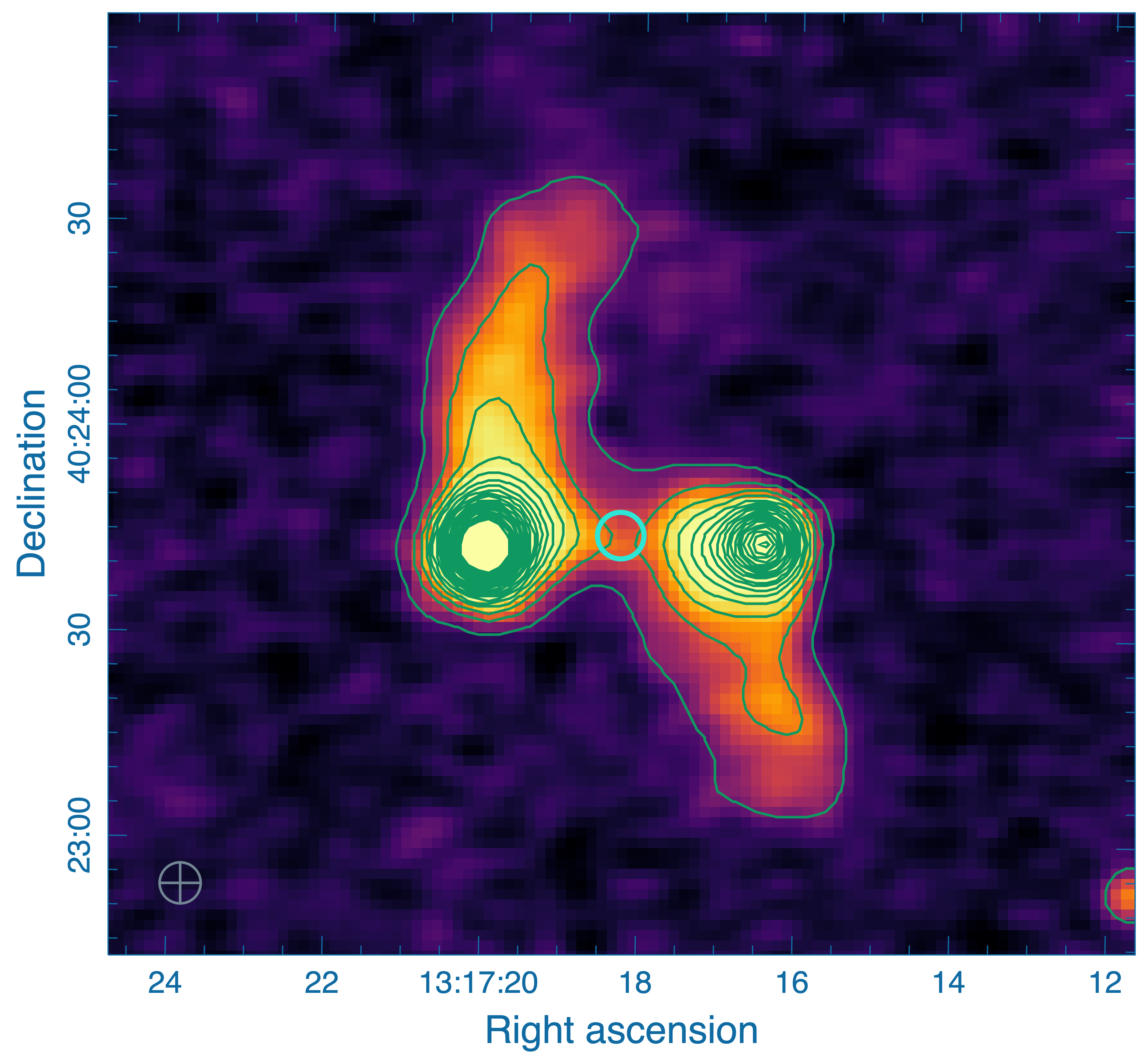}}}
\vbox{
\centerline{
\includegraphics[angle=0,width=7.0cm,height=7cm,trim={0 0cm 0 0},clip]{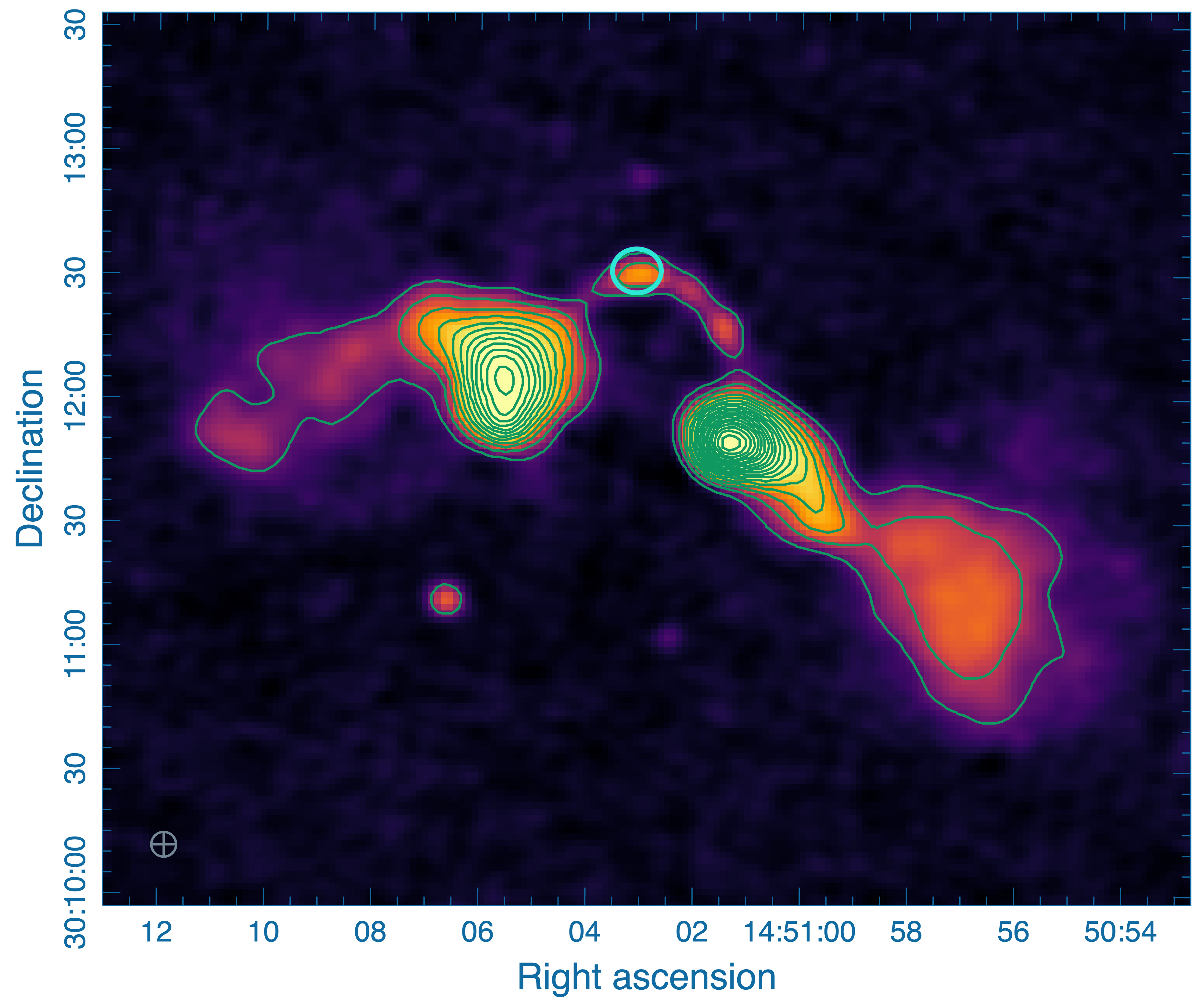}
\includegraphics[angle=0,width=7.0cm,height=7.0cm,trim={0 0cm 0 0},clip]{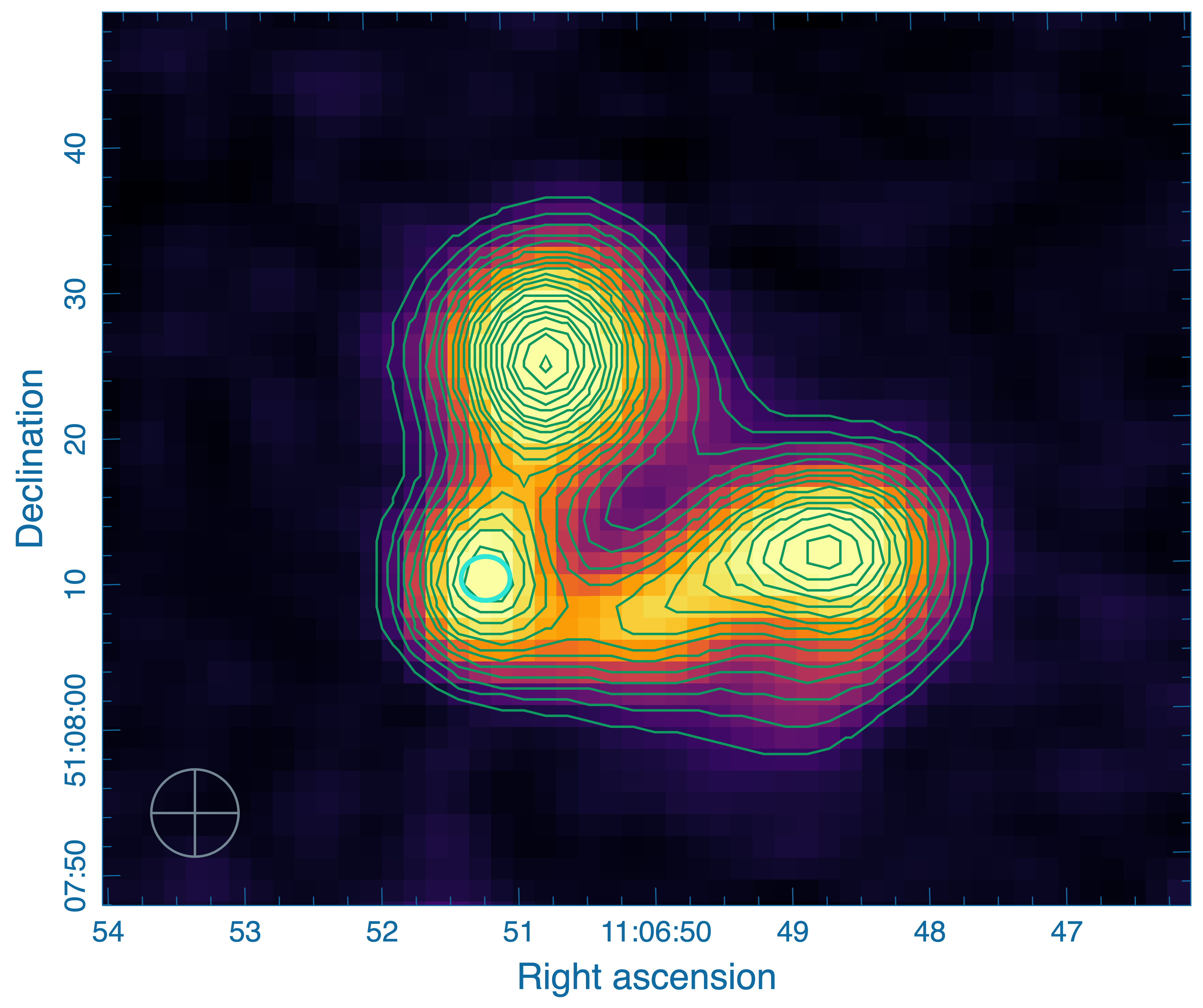}}}
\caption{Representative examples of irregular radio galaxies identified from the LoTSS DR2 survey. 
The upper panels show examples of winged radio galaxies, including X-shaped and Z-shaped morphologies, where faint off-axis emission is oriented at an angle to the main radio axis. 
The lower panels show bent-tail radio galaxies, including wide-angle tail (WAT) and narrow-angle tail (NAT) sources, in which the radio jets or lobes are bent away from a straight structure. 
These morphologies illustrate how radio jets and lobes can be shaped by jet reorientation, plasma backflow, host-galaxy motion, and interaction with the surrounding medium.}
\label{fig:irregular_examples}
\end{figure*}

Winged radio galaxies (WRGs) are another important class of irregular radio morphology, in which the radio emission is not confined to a single straight jet--lobe axis. These sources show a pair of bright, active radio lobes together with fainter, low-surface-brightness extensions, or ``wings'', oriented at a significant angle to the main radio axis, producing an `X'-shaped or butterfly-like structure in some sources \citep{Leahy1992, Cheung2007, Yang2019, Bera2020}. Based on the relative location and orientation of the wing-like emission, WRGs are commonly described as X-shaped radio galaxies (XRGs) or Z-shaped radio galaxies (ZRGs). In XRGs, the diffuse wings generally extend from the inner region of the main radio structure, whereas in ZRGs the off-axis emission produces a more laterally displaced, Z-like morphology. The physical origin of these sources remains debated, and several mechanisms have been proposed, including hydrodynamic backflow of plasma from the main lobes, jet reorientation following galaxy or black-hole mergers, precession of the jet axis, and interactions between the radio plasma and an asymmetric gaseous environment \citep{Merritt2002, GopalKrishna2012, Bera2020}. WRGs are therefore scientifically important because they provide observational clues to the history of jet activity, possible merger events, AGN duty cycles, and the role of the surrounding medium in shaping radio structures. Recent wide-area and low-frequency surveys, particularly Faint Images of the Radio Sky at Twenty Centimeters (FIRST) and LOFAR Two-metre Sky Survey (LoTSS), have greatly expanded the known samples of winged sources and have opened the way for statistical studies of their morphology, host galaxies, and environments \citep{Cheung2007, Yang2019, Bera2020, Bera2025}. Table~\ref{tab:survey_irregular} summarizes representative wide-area radio continuum surveys and published samples of bent-tail and winged radio galaxies. These samples demonstrate the rapid growth of irregular radio-galaxy catalogues in current surveys and provide the empirical basis for developing classification methods for future SKA continuum data. Figure~\ref{fig:irregular_examples} presents representative examples of irregular radio galaxies from the LoTSS DR2 survey, including winged sources with X-shaped and Z-shaped structures and bent-tail sources with WAT and NAT morphologies, illustrating how radio jets and lobes can be distorted by jet reorientation, plasma backflow, host-galaxy motion, and interaction with the surrounding medium.

The increasing depth and angular resolution of modern radio continuum surveys have transformed the study of irregular radio galaxies from the identification of individual unusual objects into a population-based field. Wide-area surveys such as the VLA FIRST, the LOFAR LoTSS,  Rapid ASKAP Continuum Survey (RACS), the Evolutionary Map of the Universe (EMU) pilot survey, MeerKAT International GHz Tiered Extragalactic Exploration (MIGHTEE), and the VLA Sky Survey (VLASS) have revealed large samples of extended and complex radio sources across large areas of the sky \citep{Becker1995, Shimwell2019, Shimwell2022, McConnell2020, Hale2021, Heywood2022, Lacy2020}. These surveys provide the observational foundation for developing visual and automated classification methods, measuring radio morphology, and connecting radio structures with host-galaxy and environmental properties. However, many faint wings, diffuse tails, remnant lobes, and low-surface-brightness features remain difficult to detect or classify in existing data, particularly for distant or low-luminosity systems.

\begin{table*}[ht]
\centering
\caption{\textbf{Major radio continuum surveys and published counts of bent-tail and winged radio sources.}
The numbers listed in the table are total published counts from the cited catalogues and should not be summed across different surveys or studies, because the samples differ in observing frequency, sky coverage, angular resolution, sensitivity, source-selection criteria, and classification method. 
For LoTSS, the bent-tail count combines 55 head--tail sources identified in LoTSS DR1 by \citet{Pal2023} and 5024 bent-tail sources identified in LoTSS DR2 and the winged or X-shaped count is from \citet{Bera2025}. 
For TGSS ADR1, the bent-tail and winged/X-shaped counts are from \citet{Bhukta2022}. 
For FIRST, the manual bent-tail count combines the published catalogues of \citet{Missaglia2019}, \citet{Pan2021}, and \citet{Sasmal2022}, while the ML count is from the machine-learning plus visual-inspection catalogue of \citet{Lao2025}. 
The FIRST winged/X-shaped count combines the catalogues of \citet{Yang2019} and \citet{Bera2020}. 
Manual denotes manual-visual classification, and ML denotes machine-learning-based selection followed by validation.}
\label{tab:survey_irregular}
\scriptsize
\setlength{\tabcolsep}{3.2pt}
\begin{tabular}{lcccccccl}
\hline
\textbf{Survey} & \textbf{Freq.} & \textbf{Res.} & \textbf{Sens.} & \textbf{Sky} & \textbf{Sources} & \textbf{Bent-tail} & \textbf{Winged/X} & \textbf{Key refs.} \\
 & (MHz) & ($''$) & (mJy/bm) & (deg$^{2}$) & (million) & (Manual/ML) & (Manual) &  \\
\hline
LoTSS & 144 & 6 & 0.095 & 5,634 & 4.4 & 5,079 / --- & 1,024 & \citet{Shimwell2022, Pal2023}\\
&&&&&&&&\citet{Bera2025} \\
TGSS ADR1 & 150 & 25 & 2--5 & 36,900 & 0.62 & 264 / --- & 58 & \citet{Intema2017, Bhukta2022} \\
FIRST & 1,400 & 5 & 0.13 & 10,575 & 0.9 & 1,173 / 4,876 & 586 & \citet{Becker1995, Missaglia2019, Pan2021}\\
&&&&&&&&\citet{Sasmal2022, Lao2025, Yang2019}\\
&&&&&&&& \citet{Bera2020} \\
NVSS & 1,435 & 45 & 0.45 & 33,800 & 2.0 & --- / --- & $\sim$100 & \citet{Cheung2007} \\
VLASS & 3,000 & 2.5 & 0.07 & 33,885 & 5.3 & --- / --- & --- & \citet{Lacy2020} \\
RACS & 887--1655 & 15 & 0.25 & 36,656 & 4.0 & --- / --- & --- & \citet{McConnell2020} \\
\hline
\end{tabular}
\end{table*}

The Square Kilometre Array Observatory (SKAO) will provide a major advance for studies of irregular radio galaxies through the SKA-Low and SKA-Mid telescopes. In the context of the updated Advancing Astrophysics with the SKA science book, the primary focus is the Array Assembly 4 (AA4) design baseline, while the earlier AA* staged-delivery configuration should also be considered when assessing which parts of this science may be achievable before the full AA4 capability. The improved sensitivity, angular resolution, frequency coverage, and imaging fidelity expected from SKA-Low and SKA-Mid will be particularly important for detecting faint extended emission and resolving the detailed structures of radio jets, lobes, tails, and wings. Such data would allow irregular radio galaxies to be identified over much larger samples and across wider ranges of redshift, radio power, host-galaxy type, and environment than is currently possible.

The scientific importance of these irregular sources extends beyond morphological classification. BT radio sources can be used to probe ram-pressure bending, galaxy motion, and the dynamical state of groups and clusters. In contrast, winged radio galaxies provide clues to jet reorientation, plasma backflow, merger history, episodic AGN activity, and the influence of the surrounding medium \citep{Begelman1979, Dehghan2014, Mguda2015, GopalKrishna2012, Bera2020, Hardcastle2020}. By combining radio continuum data with optical, infrared, and X-ray surveys, future SKA studies would make it possible to investigate how radio morphology depends on host-galaxy properties, AGN duty cycle, jet power, cluster environment, and large-scale structure. This chapter, therefore, focuses on the identification, classification, and physical interpretation of bent-tail, winged, and other non-standard radio galaxies as a route to understanding jet--environment interactions and radio-mode feedback in the SKA era.

The proposed approach for future SKA studies should combine community-established methods with new automated techniques. Visual inspection remains essential for rare and complex radio sources, especially when diffuse emission, blended components, or ambiguous host associations make automated classification uncertain. At the same time, machine-learning approaches can improve the efficiency of candidate selection in large-area surveys, provided that they are trained and validated using representative samples from existing radio data \citep{Alhassan2018, Galvin2020, Mostert2021, Alegre2022, Barkus2022, Lao2025}. A hybrid framework that combines automated detection, quantitative morphology measurements, host-galaxy identification, and expert validation offers a practical route for building reliable samples of bent-tail, winged, and other irregular radio galaxies from future SKA continuum data.

\section{Data and Methodology}
The identification and interpretation of irregular radio galaxies require a workflow that connects radio morphology with host-galaxy and environmental information. In the SKA era, large image volumes and many faint extended sources will make it necessary to combine automated methods with expert validation. The methodology adopted here follows three connected steps: identifying irregular radio morphologies, associating the radio emission with the correct host galaxy, and characterizing the surrounding environment. These steps are essential because the physical interpretation of bent-tail and winged radio galaxies depends not only on their radio structures, but also on their host properties, redshifts, and large-scale environments.

The following subsections describe the identification of irregular radio sources and host-galaxy association methods. The environmental analysis is discussed separately in Section~4.

\subsection{Machine-learning methods for identifying irregular radio sources}

The identification of irregular radio sources in large continuum surveys requires methods that are both efficient and scientifically reliable. Traditional visual inspection remains valuable because human classifiers can recognise complex, diffuse, or ambiguous radio structures that are difficult to describe with simple catalogue parameters. However, the expected source density and image volume from future SKA continuum surveys will make purely manual classification impractical. Machine-learning methods, therefore, provide an important route for selecting candidate irregular radio galaxies, measuring their morphology, and prioritising sources for expert validation.

Several machine-learning approaches have already been developed for radio-source classification and cross-identification. Supervised Convolutional Neural Networks (CNNs), which learn image features directly from labelled radio cutouts, have been used to separate compact and extended radio sources and to classify broad radio-morphology classes \citep{Alhassan2018}. Unsupervised methods, such as self-organising maps, have also been applied to radio images to group sources with similar morphologies without requiring predefined class labels, making them useful for finding rare or unusual structures \citep{Galvin2020, Mostert2021}. Other approaches focus on the association of multi-component radio emission with the correct host galaxy, including machine-learning assisted cross-matching and ridgeline-based methods that trace the extended radio structure of complex sources \citep{Alegre2022, Barkus2022}. These studies show that machine learning is most effective when it is combined with astrophysical validation, rather than used as a fully automatic replacement for expert classification.

For the specific problem of identifying bent-tail, winged, and other non-standard radio galaxies, image-segmentation methods are particularly useful. In this context, segmentation means identifying the pixels or regions of an image that belong to a radio source, instead of assigning only a single class label to the whole image. Lao et al. developed the Radio Galaxy Classification with Mask Transformer (RGCMT) framework, in which a region-based convolutional neural network first detects candidate radio components and a transformer-based refinement stage improves the segmentation of extended or low-surface-brightness emission \citep{Lao2023, Lao2025}. A region-based convolutional neural network is a deep-learning model that first proposes possible source regions in an image and then classifies and segments them. A transformer-based refinement stage uses attention mechanisms to improve the shape and boundary information of the detected source masks. This type of approach is well-suited to radio galaxies because the scientific information is contained not only in the presence of a source, but also in the geometry of its jets, lobes, tails, and diffuse extensions.

For future SKA studies, a practical classification strategy should therefore be hybrid. First, automated methods can be used to identify candidate irregular sources from large-area radio images. Second, quantitative morphological parameters, such as angular size, jet opening angle, lobe asymmetry, curvature, and surface-brightness distribution, can be measured from the detected radio emission. Third, uncertain cases should be checked through visual inspection using radio contours, multi-frequency radio images, and optical or infrared overlays. This validation step is essential for separating genuine irregular radio galaxies from imaging artefacts, blended sources, unrelated neighbouring components, or projection effects.

The extension of these methods to SKA data will require careful validation because SKA-Low and SKA-Mid observations are expected to detect fainter and more diffuse emission than many current surveys. Training samples based only on existing data may not fully represent the range of low-surface-brightness and high-redshift morphologies that SKA observations could reveal. Therefore, transfer learning, active learning, and simulations of SKA-like images should be used to adapt existing classifiers to the depth, resolution, and noise properties of SKA continuum data. In this framework, machine learning would provide a scalable candidate-selection and measurement tool, while expert inspection and multi-wavelength information would ensure the reliability needed for physical studies of jet--environment interactions and radio-galaxy evolution.

\subsection{Host galaxy identification}

Reliable host-galaxy identification is essential for connecting radio morphology with the physical properties of the galaxy that launches the jets. For compact radio sources, the radio centroid or radio core usually lies close to the optical or infrared counterpart of the host galaxy. However, host identification becomes more challenging for extended and irregular radio galaxies, where the radio emission may be separated into multiple components, the core may be weak or undetected, and the radio lobes or tails may extend far beyond the stellar body of the host. In such cases, a simple positional match can produce incorrect associations, and a combination of statistical, visual, and multi-wavelength methods is required.

For relatively compact or core-detected sources, a nearest-neighbour association provides a useful first step. In this approach, the radio position is matched to the closest optical or infrared source within a defined search radius, taking into account the positional uncertainties of the radio and ancillary catalogues \citep{deRuiter1977}. Optical and infrared catalogues such as the Sloan Digital Sky Survey (SDSS), Dark Energy Spectroscopic Instrument Legacy Imaging Surveys (DESI), Panoramic Survey Telescope and Rapid Response System (Pan-STARRS), All-sky Wide-field Infrared Survey Explorer (AllWISE), and unblurred Wide-field Infrared Survey Explorer (unWISE) provide the photometric information required to identify the likely host galaxies and to estimate their redshifts, stellar masses, colours, and other physical properties \citep{York2000, Dey2019, Chambers2016, Wright2010, Lang2014}. For irregular radio galaxies, these host properties are necessary for measuring projected linear sizes, radio luminosities, and the dependence of morphology on galaxy type and environment.

A more robust method for host identification is the likelihood-ratio (LR) technique, which estimates the probability that a given optical or infrared source is the true counterpart of a radio source. The LR method combines the positional offset between the radio and optical/infrared source, the expected positional uncertainty, the magnitude distribution of genuine counterparts, and the surface density of background objects \citep{Sutherland1992, McAlpine2012}. It can be written as
\begin{equation}
LR = \frac{q(m) f(r)}{n(m)},
\end{equation}
where $q(m)$ is the expected magnitude distribution of true counterparts, $f(r)$ is the probability distribution of positional offsets, and $n(m)$ is the surface density of unrelated background sources with magnitude $m$. This method has been used effectively in modern radio surveys, including the LoTSS, where optical and infrared identifications were obtained using a combination of LR matching, automated association methods, and visual inspection \citep{Williams2019, Kondapally2021, Hardcastle2023}.

Visual inspection remains essential for extended, multi-component, and morphologically complex radio galaxies. For these sources, the correct host is often identified by examining radio contours overlaid on optical or infrared images, locating the symmetry centre of the radio structure, and checking whether a plausible galaxy lies between the lobes, jets, tails, or wing-like emission. Citizen-science and expert-inspection projects such as Radio Galaxy Zoo and LOFAR Galaxy Zoo have demonstrated the importance of visual association for complex radio sources and have provided training samples for automated host-identification methods \citep{Banfield2015, Williams2019, Alger2018, Hardcastle2023}. Therefore, visual inspection should not be treated only as a final check, but as an important component of the identification process for sources where the radio morphology is not well represented by catalogue positions alone.

For future SKA continuum data, host-galaxy identification should follow a hybrid strategy. Compact and simple sources can be associated using nearest-neighbour or LR-based methods, while extended and irregular sources should be processed through algorithms that first group related radio components and then assign the most probable host galaxy. Machine-learning approaches may assist this process by recognising multi-component radio structures and ranking possible host candidates, but their results must be calibrated with visually confirmed samples. Each host association should therefore include a reliability flag, indicating whether the identification was obtained through nearest-neighbour matching, LR analysis, automated multi-component association, visual inspection, or a combination of these methods. Such a framework would provide the reliable host identifications required for studying the connection between radio morphology, AGN activity, host-galaxy properties, and environment in the SKA era.

\section{Role of SKA Continuum Surveys}

The SKA continuum surveys will provide a powerful opportunity to study irregular radio galaxies with much improved sensitivity, angular resolution, and image quality. In this chapter, the discussion is framed around the AA4 baseline, with AA* representing an earlier stage that may already enable important preparatory studies. The relevant technical details for AA4 and AA* should be taken from the official SKAO technical documentation and sensitivity tools.

The main advantage of SKA continuum observations for this science case is the ability to detect and resolve faint extended radio emission. Irregular radio galaxies often contain low-surface-brightness structures, such as bent jets, diffuse tails, weak bridges, asymmetric lobes, and wing-like extensions. These features can be difficult to identify in present surveys, particularly when the sources are faint, distant, or located in complex environments. The improved sensitivity and imaging fidelity of SKA-Low and SKA-Mid will make it possible to recover such structures more reliably and to study their connection with host galaxies and surrounding environments.

Current surveys such as FIRST, LoTSS, RACS, EMU, MIGHTEE, and VLASS have already shown the importance of wide-area radio imaging for discovering complex radio morphologies \citep{Becker1995, Shimwell2019, Shimwell2022, McConnell2020, Norris2021, Heywood2022, Lacy2020}. These surveys provide the foundation for developing classification methods, host-identification techniques, and statistical studies of radio morphology. The SKA will extend this work by detecting fainter radio plasma, resolving finer jet and lobe structures, and increasing the number of sources suitable for population studies.

The scientific role of SKA continuum surveys is therefore not limited to finding more irregular radio galaxies. These data will allow measurements of bending angle, angular size, projected linear size, lobe asymmetry, radio luminosity, spectral index, surface brightness, and polarization properties. When combined with optical, infrared, and X-ray information, these measurements can be used to investigate how radio morphology depends on host-galaxy properties, AGN activity, jet power, and the surrounding group or cluster environment \citep{Croston2019, Hardcastle2020}. In this way, SKA continuum surveys will provide a direct route to studying jet--environment interaction and radio-mode feedback.
The main physical and observational parameters relevant for irregular radio-galaxy studies are summarized in Appendix~\ref{app:irregular_params}.

\subsection{Expected source yields}

The number of irregular radio galaxies expected from SKA continuum surveys will depend on the adopted observing strategy, including survey area, observing frequency, angular resolution, sensitivity, and the surface-brightness limit. It will also depend on the selection criteria used to define bent-tail, winged, and other non-standard radio galaxies. For this reason, source-yield estimates should be presented as approximate expectations rather than fixed predictions.

A useful approach is to use existing surveys as empirical guides. FIRST, LoTSS, RACS, EMU, MIGHTEE, and VLASS provide information on the surface density of extended radio sources and on the fraction of sources that show irregular morphologies \citep{Becker1995, Shimwell2022, McConnell2020, Norris2021, Heywood2022, Lacy2020}. These observed fractions can be combined with SKA survey depth and sky coverage to estimate the likely number of candidate irregular radio galaxies. Such estimates should also consider completeness and reliability, because faint diffuse structures may be missed, while unrelated neighbouring sources or imaging artefacts may sometimes mimic irregular morphology.

Bent-tail radio galaxies are expected to form an important population in SKA continuum surveys because their distorted jets and lobes trace the interaction between radio plasma and dense environments. Their morphology is commonly linked to ram pressure, galaxy motion, and the dynamical state of groups and clusters \citep{Begelman1979, Dehghan2014, Mguda2015}. Winged radio galaxies are also expected to benefit from the improved surface-brightness sensitivity of the SKA, because their off-axis emission is often faint and diffuse. Larger and deeper samples of these sources will make it possible to test models involving plasma backflow, jet reorientation, merger activity, and environmental effects \citep{Cheung2007, Yang2019, Bera2020}.

AA* observations may already be useful for testing source-finding methods, host-identification procedures, and machine-learning classification pipelines on early SKA data. The AA4 baseline will then provide the sensitivity and statistical power required for larger and more complete samples. Together, these stages will help build a reliable framework for identifying irregular radio galaxies and studying their physical origin.

\subsection{Technical advantages}

Several technical strengths of SKA continuum surveys are especially important for irregular radio-galaxy studies. The first is high surface-brightness sensitivity. Many irregular sources contain faint diffuse emission that traces older synchrotron plasma, remnant lobes, or weak extensions produced by jet--environment interaction. SKA-Low will be particularly valuable for detecting steep-spectrum and aged radio plasma, while SKA-Mid will provide complementary information at higher frequencies.

The second advantage is improved angular resolution and image fidelity. Irregular morphology is identified from the shape and geometry of the radio structure, including jet curvature, lobe asymmetry, wing orientation, and the position of the host galaxy relative to the radio emission. High-quality imaging is therefore essential for distinguishing genuine physical structures from artefacts, projection effects, or unrelated neighbouring components.

The third advantage is broad frequency coverage. Multi-frequency radio data will allow spectral-index and spectral-curvature studies across different parts of a source. These measurements can help separate active jets and hotspots from older plasma and can provide information on synchrotron ageing, restarted activity, and the history of particle acceleration.

The fourth advantage is polarization capability. Polarization and rotation-measure information can trace magnetic-field structure and the magnetised plasma surrounding radio galaxies. These measurements are important for understanding how radio jets interact with the intracluster medium and how the surrounding environment influences the observed radio morphology \citep{Beck2021}.

Finally, the full scientific value of SKA continuum surveys will come from combining radio data with optical, infrared, and X-ray surveys. Host-galaxy identifications, redshifts, stellar masses, cluster properties, and environmental measurements will allow irregular radio sources to be studied as physical systems rather than only as morphological classes. This multi-wavelength approach will make it possible to connect radio morphology with AGN duty cycle, jet power, galaxy evolution, and feedback in groups and clusters.

\section{Environmental Studies and Cluster Connection}

The environments of irregular radio galaxies are central to understanding why their radio jets and lobes deviate from a straight morphology. In bent-tail radio galaxies, the curvature of the jets is commonly interpreted as the result of interaction between the radio plasma and the surrounding intracluster or intragroup medium. The motion of the host galaxy through a dense medium, together with bulk gas motions produced by cluster mergers or turbulence, can generate ram pressure that bends the jets and tails \citep{Ryle1968, Begelman1979, Blanton2000, Dehghan2014, Mguda2015}. Therefore, bent-tail sources are not only examples of unusual radio morphology, but also useful probes of dense environments, cluster dynamics, and the physical conditions of the intracluster medium (ICM).

The connection between radio morphology and environment is also important for understanding AGN feedback. Radio jets transport energy from the central active galactic nucleus into the surrounding gas, where they can inflate lobes and cavities, redistribute cosmic rays and magnetic fields, and influence the thermal evolution of groups and clusters \citep{Croston2019, Hardcastle2020}. Distorted radio morphologies provide direct evidence that this feedback process is not occurring in isolation. Instead, the shape of the radio source records the combined effects of jet power, host-galaxy motion, ICM density, pressure gradients, turbulence, and merger-driven gas flows. Studying these sources therefore provides a way to connect small-scale AGN activity with the large-scale evolution of galaxies, groups, and clusters.

Winged and other asymmetric radio galaxies also provide environmental diagnostics, although their physical interpretation can be more complex. In some systems, off-axis or wing-like emission may be related to plasma backflow, jet reorientation, restarted activity, or interaction with an asymmetric external medium \citep{Merritt2002, GopalKrishna2012, Bera2020}. Environmental information is therefore essential for distinguishing between models. For example, a source located in a disturbed cluster may favour scenarios involving ICM motions or asymmetric pressure gradients, while evidence of a merger or double nucleus in the host galaxy may support jet reorientation or restarted AGN activity. Multi-wavelength environmental studies are thus required to interpret irregular radio morphology physically, rather than treating it only as a visual classification.

A practical environmental analysis should combine radio morphology with optical, infrared, and X-ray information. Optical and infrared surveys provide host-galaxy identifications, photometric or spectroscopic redshifts, stellar masses, and galaxy-density estimates, while X-ray and Sunyaev--Zel'dovich catalogues trace the hot gas and massive cluster population. Cluster catalogues such as WHL, redMaPPer, Planck, and SPT have been widely used to identify galaxy clusters and to estimate properties such as richness, mass, and cluster radius \citep{Wen2015, Rozo2015, Bleem2015, Planck2016}. Cross-matching irregular radio galaxies with such catalogues can reveal whether bent or distorted morphologies preferentially occur in rich clusters, merging systems, cluster outskirts, or lower-mass group environments.

Redshift information is an important part of this analysis, but it should be treated with care. Not all radio galaxies detected in future SKA continuum surveys will have spectroscopic redshifts, especially in wide-area surveys. Many sources will instead rely on photometric redshifts from optical and infrared catalogues, while some faint or high-redshift hosts may not have reliable redshift estimates at all. Therefore, environmental studies should use a tiered approach. Sources with spectroscopic redshifts can be used for the most secure host--cluster associations. Sources with good photometric redshifts can be matched probabilistically to clusters using redshift uncertainty intervals rather than a single fixed threshold. Sources without reliable redshifts can still be used for angular clustering, surface-density studies, and for selecting targets for deeper optical, infrared, or spectroscopic follow-up.

For sources with reliable redshift information, cluster association can be tested using both projected separation and redshift consistency. A radio galaxy may be considered a likely cluster member when its host lies within a physically motivated projected radius, such as a fraction of $R_{200}$ or $R_{500}$, and when its redshift is consistent with the cluster redshift within the uncertainty of the measurement. The adopted redshift interval should depend on whether the redshift is spectroscopic or photometric. Spectroscopic samples allow tighter criteria, while photometric samples require broader probability-based matching. This approach avoids over-interpreting uncertain associations and allows completeness and contamination to be quantified.

The main scientific goal of such environmental studies is to determine how radio morphology depends on the environment. For bent-tail sources, one can examine how bending angle, tail length, radio luminosity, spectral index, and host-galaxy position vary with cluster mass, richness, projected cluster-centric distance, and dynamical state. These measurements can test whether strongly bent sources preferentially trace dense ICM conditions, cluster mergers, or galaxies moving rapidly through the cluster potential. For winged and asymmetric sources, the same environmental information can help determine whether the observed morphology is more closely linked to host-galaxy merger history, jet-axis changes, or external gas asymmetry.

SKA continuum data will be especially valuable because they will improve the detection of faint tails, wings, and diffuse lobes that are often missed in current surveys. When combined with multi-wavelength catalogues, these data will allow large statistical samples of irregular radio galaxies to be used as probes of AGN feedback and cluster environments. This will make it possible to move from individual case studies to population-level tests of how radio jets respond to their surroundings and how AGN feedback operates across different environments and cosmic epochs.

\section{Expected Outcomes and Science Prospects}

SKA continuum surveys will provide a major opportunity to study irregular radio galaxies as physical probes of AGN activity, jet propagation, and environment. The main outcome will not only be the discovery of larger samples of bent-tail, winged, and other non-standard radio galaxies, but also the construction of statistically useful catalogues with reliable morphology, host-galaxy information, and environmental measurements. Such catalogues will allow irregular radio galaxies to be studied as part of the broader radio-galaxy population, rather than as isolated peculiar objects.

One of the key science goals is to understand which physical conditions produce complex and asymmetric radio morphologies. For bent-tail radio galaxies, the bending of jets and lobes can be related to the motion of the host galaxy through the surrounding medium, ram pressure, and bulk gas motions in galaxy groups and clusters \citep{Begelman1979, Dehghan2014, Mguda2015}. With large SKA-selected samples, it will be possible to compare bending angle, tail length, radio luminosity, spectral index, and host-galaxy position with cluster-centric distance, richness, mass, and dynamical state. This will help determine whether strongly bent sources preferentially trace dense intracluster medium (ICM), merging clusters, or galaxies moving rapidly through the group or cluster potential.

For winged radio galaxies, SKA observations will help test whether the secondary, low-surface-brightness structures are mainly produced by hydrodynamic backflow, jet reorientation, restarted activity, or interaction with an asymmetric external medium \citep{Merritt2002, GopalKrishna2012, Bera2020}. Deep radio imaging will be especially important because the wing-like emission is often faint and may be missed in shallower surveys. Multi-frequency SKA data will allow spectral-index and spectral-ageing measurements across the active lobes and the wings. If the wings contain older plasma than the main lobes, this may support scenarios involving relic emission or jet reorientation. If the wings show a close connection with the surrounding gas distribution, this may favour backflow or environmental-deflection models.

A second major science goal is to clarify how AGN jets transfer energy to their surroundings. Radio jets inflate lobes and cavities, drive weak shocks, stir turbulence, and transport relativistic particles and magnetic fields into the circumgalactic and intracluster medium \citep{Croston2019, Hardcastle2020}. Recent hydrodynamic and magnetohydrodynamic simulations show that jet power, ambient density, cluster weather, and lobe disruption strongly influence how feedback energy is deposited into the surrounding gas \citep{Mukherjee2018, English2019, Bourne2021}. SKA observations can provide the radio constraints needed to test these models, including lobe size, surface-brightness distribution, spectral ageing, and the presence of remnant or restarted emission.

Low-frequency SKA observations will be particularly important for studying aged synchrotron plasma and the life cycle of radio galaxies. Simulations of cosmic-ray electron transport show that radio galaxies can seed the ICM with relativistic electrons and magnetic fields, producing fossil electron populations that may survive and be re-accelerated by shocks or turbulence over long timescales \citep{Vazza2021, Vazza2023}. Observationally, this means that faint diffuse tails, remnant lobes, and cluster-scale radio emission can be used to trace the past activity of radio galaxies and the long-term impact of AGN feedback on the surrounding medium.

Polarization and rotation-measure studies will provide another important diagnostic. The polarization structure of radio lobes and tails traces magnetic-field ordering, while rotation measure probes the magnetised plasma along the line of sight. Recent simulations show that Faraday-rotation information can help separate the effects of intrinsic jet power from the effects of environmental density and magnetic-field strength \citep{Jerrim2024}. With SKA polarization data, bent-tail and winged radio galaxies can therefore be used to study both the internal magnetic structure of radio plasma and the magneto-ionic properties of the surrounding ICM.

The combination of SKA radio data with optical, infrared, and X-ray surveys will also make it possible to connect radio morphology with host-galaxy evolution. Host identifications and redshift estimates will allow measurements of radio luminosity, projected linear size, stellar mass, colour, and environment. These quantities can be used to test whether irregular morphologies are more common in particular host-galaxy populations, at particular stages of AGN activity, or in particular large-scale environments. For systems with evidence of galaxy mergers or disturbed hosts, simulations of jet feedback in merging galaxies provide a useful framework for interpreting jet reorientation, recurrent activity, and the effect of jets on the circumgalactic medium \citep{Talbot2024}.

Overall, the expected scientific return of SKA continuum surveys will be a more complete physical picture of irregular radio galaxies. Large samples will allow population-level tests of how jet morphology depends on environment, while deep multi-frequency and polarization data will reveal the plasma history within individual sources. By linking observations with modern simulations, SKA studies will provide new constraints on jet--environment coupling, radio-mode feedback, the duty cycle of AGN activity, and the role of radio galaxies in the evolution of groups and clusters.

\section{Conclusions}

Irregular radio galaxies provide important laboratories for studying the interaction between radio jets, host galaxies, and their surrounding environments. In this chapter, the term ``irregular radio morphology'' refers to radio sources in which the jets and lobes deviate from a straight, symmetric structure. Bent-tail and winged radio galaxies are two important examples of such systems. Their distorted radio structures contain information about jet propagation, galaxy motion, intracluster-medium conditions, episodic AGN activity, and radio-mode feedback \citep{Begelman1979, Dehghan2014, Mguda2015, GopalKrishna2012, Bera2020, Hardcastle2020}.

The SKA continuum surveys will greatly improve the study of these sources by providing higher sensitivity, better image fidelity, broad frequency coverage, and polarimetric capability. SKA-Low will be especially useful for detecting faint, steep-spectrum, and aged synchrotron plasma, while SKA-Mid will provide the angular resolution needed to resolve the detailed structures of jets, lobes, tails, and wings. In the AA* stage, early SKA observations can already be used to test classification methods, host-identification procedures, and environmental-analysis pipelines. The AA4 baseline will then provide the sensitivity and statistical power required for larger and more complete samples of irregular radio galaxies.

A reliable study of irregular radio galaxies in the SKA era will require a hybrid approach. Machine-learning methods can efficiently identify candidate sources in large radio images, measure morphological parameters, and prioritize unusual objects for further inspection. However, visual validation will remain essential for complex, diffuse, or multi-component sources, where automated methods can be affected by blending, imaging artefacts, or uncertain host associations \citep{Alhassan2018, Galvin2020, Mostert2021, Alegre2022, Barkus2022, Lao2023, Lao2025}. Host-galaxy identification should also combine nearest-neighbour matching, likelihood-ratio methods, multi-component association, and expert inspection, particularly for extended sources without a clearly detected radio core \citep{Williams2019, Kondapally2021, Hardcastle2023}.

The scientific return of SKA studies will come from combining radio morphology with multi-wavelength information. Optical and infrared data will provide host identifications, redshifts, stellar masses, and galaxy properties, while X-ray and cluster catalogues will trace the hot gas and large-scale environment. This combination will make it possible to test how jet bending, wing formation, lobe asymmetry, radio luminosity, spectral index, and polarization depend on host-galaxy properties, AGN duty cycle, cluster mass, and dynamical state. Such studies will connect the observed morphology of radio galaxies with the physical processes that regulate galaxy and cluster evolution.

The interpretation of SKA observations will also benefit from comparison with modern simulations of AGN jets and feedback. Hydrodynamic and magnetohydrodynamic simulations can help determine how jet power, ambient density, cluster weather, magnetic fields, and cosmic-ray transport shape radio structures and distribute feedback energy into the surrounding medium \citep{Mukherjee2018, English2019, Bourne2021, Vazza2021, Vazza2023, Jerrim2024}. By linking large observational samples with theoretical modelling, SKA continuum surveys will provide a powerful framework for understanding jet--environment coupling, the life cycle of radio galaxies, and the role of radio-mode feedback in groups and clusters across cosmic time.

\appendix

\section{Physical and Observational Parameters}
\label{app:irregular_params}

Table~\ref{tab:irregular_params} summarizes the main physical and observational quantities that can be measured for irregular radio galaxies in SKA continuum studies. These parameters are relevant for bent-tail, winged, and other non-standard radio galaxies whose jets or lobes deviate from a straight, symmetric structure.

\begin{table*}[ht]
\centering
\caption{\textbf{Primary physical and observational parameters for irregular radio-galaxy studies with SKA continuum data.}
These quantities are intended for sources whose jets and lobes deviate from a straight, symmetric structure, including bent-tail and winged radio galaxies. The accuracy of each parameter will depend on image quality, angular resolution, source brightness, redshift availability, and the reliability of the host-galaxy association.}
\label{tab:irregular_params}
\scriptsize
\setlength{\tabcolsep}{4pt}
\renewcommand{\arraystretch}{1.15}
\begin{tabular}{p{3.2cm} c p{10.0cm}}
\hline
\textbf{Parameter} & \textbf{Units} & \textbf{Description / Estimation Method} \\
\hline

Right Ascension (RA), Declination (Dec) 
& deg 
& Position of the radio core, host-galaxy centroid, or best-estimated source centre. \\

Integrated Flux Density ($S_{\nu}$) 
& mJy 
& Total radio flux density measured over the source region at a given observing frequency. \\

Peak Flux Density 
& mJy beam$^{-1}$ 
& Maximum brightness within the source region; useful for identifying compact cores or hotspots. \\

Spectral Index ($\alpha$) 
& --- 
& Estimated from multi-frequency radio flux densities using $S_{\nu}\propto\nu^{-\alpha}$; useful for identifying active and aged synchrotron plasma. \\

Angular Size 
& arcsec 
& Largest angular extent of the radio emission, measured from radio contours or source masks. \\

Projected Linear Size 
& kpc 
& Physical size derived from angular size and angular-diameter distance using the host-galaxy redshift. \\

Bending / Opening Angle ($\psi$) 
& deg 
& Projected angle between the jet or tail directions; used to quantify bent-tail morphologies such as WAT and NAT sources. \\

Wing / Off-axis Angle 
& deg 
& Orientation of faint off-axis emission relative to the main radio axis in winged, X-shaped, or Z-shaped radio galaxies. \\

Arm-length Ratio 
& --- 
& Ratio of projected distances from the host galaxy to the opposite lobes or tails; traces structural asymmetry. \\

Flux-density Ratio 
& --- 
& Ratio of integrated flux densities between opposite lobes, tails, or wings; useful for studying asymmetry, environment, and orientation. \\

Radio Luminosity ($L_{\nu}$) 
& W Hz$^{-1}$ 
& Rest-frame radio luminosity estimated from integrated flux density, redshift, and spectral-index correction. \\

Surface Brightness ($\Sigma_{\nu}$) 
& Jy arcsec$^{-2}$ 
& Mean radio brightness over the diffuse source region; important for detecting faint tails, wings, bridges, and remnant plasma. \\

Redshift ($z$) 
& --- 
& Spectroscopic or photometric redshift of the host galaxy; the redshift type and uncertainty should be recorded. \\

Host Stellar Mass ($M_{\star}$) 
& $M_{\odot}$ 
& Host-galaxy stellar mass estimated from optical/infrared photometry or spectral-energy-distribution fitting. \\

Host Identification Flag 
& --- 
& Reliability flag indicating whether the host was identified by nearest-neighbour matching, likelihood-ratio analysis, automated association, visual inspection, or a combined method. \\

Environment Density ($\Sigma_{5}$) 
& Mpc$^{-2}$ 
& Local projected galaxy density estimated from nearest-neighbour statistics around the host galaxy. \\

Cluster Association 
& --- 
& Association with an optical, X-ray, or SZ-selected galaxy group or cluster using projected separation and redshift consistency. \\

Cluster-centric Distance 
& Mpc or $R/R_{200}$ 
& Projected distance between the radio host and the cluster centre, expressed in physical units or normalized by the cluster radius. \\

Morphological Class 
& --- 
& Radio morphology class, such as FR-I, FR-II, bent-tail, WAT, NAT, winged, X-shaped, Z-shaped, or other irregular morphology. \\

Polarization Fraction 
& \% 
& Fractional polarization of the radio emission; traces magnetic-field ordering in jets, lobes, tails, and wings. \\

Rotation Measure (RM) 
& rad m$^{-2}$ 
& Faraday-rotation measurement used to probe magnetised plasma in the radio source and surrounding environment. \\

Classification Reliability 
& --- 
& Quality flag describing the confidence of the morphological classification and possible contamination from artefacts, blending, or projection effects. \\

\hline
\end{tabular}
\end{table*}

\section*{Acknowledgements}

The authors acknowledge the SKA Observatory and the broader SKA science community for providing the scientific and technical framework within which this chapter has been developed. This work has benefited from the scientific and technical framework developed by the SKA Observatory and the broader SKA science community for SKA-Low and SKA-Mid.

This chapter also makes use of knowledge and results from several major radio continuum surveys that have shaped the current understanding of extended and irregular radio galaxies, including FIRST \citep{Becker1995,Helfand2015}, LoTSS \citep{Shimwell2019,Shimwell2022}, TGSS ADR1 \citep{Intema2017}, NVSS \citep{Condon1998}, VLASS \citep{Lacy2020}, RACS \citep{McConnell2020}, EMU \citep{Norris2021}, and MIGHTEE \citep{Heywood2022}. These surveys provide the observational basis for identifying bent-tail, winged, and other complex radio morphologies, and for developing the classification and host-identification methods discussed in this chapter.

The authors acknowledge the use of multi-wavelength survey resources that are essential for host-galaxy identification and environmental studies, including SDSS \citep{York2000}, DESI Legacy Imaging Surveys \citep{Dey2019}, Pan-STARRS \citep{Chambers2016}, AllWISE \citep{Wright2010}, and unWISE \citep{Lang2014}. We also acknowledge the use of optical, infrared, X-ray, and galaxy-cluster catalogues that support environmental studies of radio galaxies, including WHL \citep{Wen2015}, redMaPPer \citep{Rozo2015}, Planck \citep{Planck2016}, and SPT cluster catalogues \citep{Bleem2015}.

The authors also acknowledge the developers of widely used astronomical software and machine-learning tools, including Astropy \citep{Astropy2018} and PyTorch \citep{Paszke2019}, which support radio-image analysis, catalogue handling, cross-matching, and automated classification. Finally, we sincerely thank the reviewers for their valuable and constructive comments, which have significantly improved the clarity, structure, scientific motivation, and overall presentation of this chapter.

\bibliographystyle{abbrvnat}
\input{journal-names.tex}
\bibliography{chapter4}

\end{document}

%% file: journal-names.tex
\newcommand{\actaa}{Acta Astron.} % Acta Astronomica
\newcommand{\araa}{ARA\&A} % Annual Review of Astron and Astrophys
\newcommand{\aar}{A\&ARv} % Astrononmy \& Astrophysics Review
\newcommand{\aapr}{A\&ARv} % Astronomy\&Astrophysics Reviews
\newcommand{\ab}{Astrobiol.} % Astrobiology
\newcommand{\aj}{AJ} % Astronomical Journal
\newcommand{\apj}{ApJ} % Astrophysical Journal
\newcommand{\apjl}{ApJL} % Astrophysical Journal, Letters
\newcommand{\apjs}{ApJSS} % Astrophysical Journal, Supplement
\newcommand{\ao}{Appl. Opt.} % Applied Optics
\newcommand{\apss}{Astro. \& Space Sci.} % Astrophysics and Space Science
\newcommand{\aap}{A\&A} % Astronomy and Astrophysics
\newcommand{\aaps}{A\&AS.} % Astronomy and Astrophysics, Supplement
\newcommand{\baas}{Bull. Am. Astron. Soc.} % Bulletin of the AAS
\newcommand{\caa}{Chinese A\&A} % Chinese Astronomy and Astrophysics
\newcommand{\cjaa}{Chinese J. A\&A} % Chinese Journal of Astronomy and Astrophysics
\newcommand{\cqg}{Class. Quantum Gravity} % Classical and Quantum Gravity
\newcommand{\gal}{Galaxies} % Galaxies
\newcommand{\gca}{Geo. Cosmo. Acta} % Geochimica Cosmochimica Acta
\newcommand{\icarus}{Icarus} % Icarus
\newcommand{\jcap}{JCAP} % Journal of Cosmology and Astroparticle Physics
\newcommand{\jgr}{J. Geophys. Res.} % Journal of Geophysics Research
\newcommand{\jgrp}{J. Geophys. Res. Planets} % Journal of Geophysics Research: Planets
\newcommand{\jqsrt}{J. Quant. Spectrosc. Radiat. Transf.} % Journal of Quantitiative Spectroscopy and Radiative Transfer
\newcommand{\memsai}{Mem. SAIt} % Mem. Societa Astronomica Italiana
\newcommand{\mnras}{MNRAS} % Monthly Notices of the RAS
\newcommand{\nat}{Nature} % Nature
\newcommand{\nastro}{Nat. Astron.} % Nature Astronomy
\newcommand{\ncomms}{Nat. Commun.} % Nature Communications
\newcommand{\nphys}{Nat. Phys.} % Nature Physics
\newcommand{\na}{New Astron.} % New Astronomy
\newcommand{\nar}{New Astron. Rev.} % New Astronomy Review
\newcommand{\physrep}{Phys. Rep.} % Physics Reports
\newcommand{\pra}{Phys. Rev. A} % Physical Review A: General Physics
\newcommand{\prb}{Phys. Rev. B} % Physical Review B: Solid State
\newcommand{\prc}{Phys. Rev. C} % Physical Review C
\newcommand{\prd}{Phys. Rev. D} % Physical Review D
\newcommand{\pre}{Phys. Rev. E} % Physical Review E
\newcommand{\prx}{Phys. Rev. X} % Physical Review X
\newcommand{\prl}{Phys. Rev. Let.} % Physical Review Letters
\newcommand{\psj}{Planet. Sci. J.} % Planetary Science Journal
\newcommand{\planss}{Planet. Space Sci.} % Planetary Space Science
\newcommand{\pnas}{Proc. Natl Acad. Sci. USA} % Proceedings of the US National Academy of Sciences
\newcommand{\procspie}{Proc. SPIE} % Proceedings of the SPIE
\newcommand{\pasa}{PASA} % Publications of the Astron.  Soc. of Australia
\newcommand{\pasj}{PASJ} % Publications of the Astron.  Soc. of Japan 
\newcommand{\pasp}{PASP} % Publications of the Astron.  Soc. of the Pacific
\newcommand{\rmxaa}{RMXAA} % Revista Mexicana de Astronomia y Astrofisica
\newcommand{\sci}{Science} % Science
\newcommand{\sciadv}{Sci. Adv.} % Science Advances
\newcommand{\solphys}{Sol. Phys.} % Solar Physics
\newcommand{\sovast}{Soviet Ast.} % Soviet Astronomy
\newcommand{\ssr}{Space Sci. Rev.} % Space Science Reviews
\newcommand{\uni}{Universe} % Universe